\documentclass[conference]{IEEEtran}
\IEEEoverridecommandlockouts
\usepackage{cite}
\usepackage{amsmath,amssymb,amsfonts}
\usepackage{algorithmic}
\usepackage{graphicx}
\usepackage{subfigure}
\usepackage{textcomp}
\usepackage{xcolor}
\usepackage{enumerate}
\usepackage{caption}
\usepackage{enumitem}
\def\BibTeX{{\rm B\kern-.05em{\sc i\kern-.025em b}\kern-.08em
    T\kern-.1667em\lower.7ex\hbox{E}\kern-.125emX}}
\begin{document}

\title{Privacy-Preserving Model Aggregation for Asynchronous Federated Learning}

\author{\IEEEauthorblockN{Jianxiang Zhao, Xiangman Li, and Jianbing Ni} \IEEEauthorblockA{Department of Electrical \& Computer Engineering, Queen's University, Kingston, Canada K7L 3N6\\}
 Email: \{jianxiang.zhao, xiangman.li, jianbing.ni\}@queensu.ca}

\maketitle

\begin{abstract}
We present a novel privacy-preserving model aggregation for asynchronous federated learning, named PPA-AFL that removes the restriction of synchronous aggregation of local model updates in federated learning, while enabling the protection of the local model updates against the server. In PPA-AFL, clients can proactive decide when to engage in the training process, and sends local model updates to the server when the updates are available. Thus, it is not necessary to keep synchronicity with other clients. To safeguard client updates and facilitate local model aggregation, we employ Paillier encryption for local update encryption and support homomorphic aggregation. Furthermore, secret sharing is utilized to enable the sharing of decryption keys and facilitate privacy-preserving asynchronous aggregation. As a result, the server remains unable to gain any information about the local updates while asynchronously aggregating to produce the global model. We demonstrate the efficacy of our proposed PPA-AFL framework through comprehensive complexity analysis and extensive experiments on a prototype implementation, highlighting its potential for practical adoption in privacy-sensitive asynchronous federated learning scenarios. \footnote{A part of the study has been published in Jianxiang Zhao's Master thesis. The authors own the copyright to the thesis as a whole and it is allowed to republish according to Intellectual Property Guidelines at Queen's University.}
\end{abstract}

\begin{IEEEkeywords}
component, formatting, style, styling, insert
\end{IEEEkeywords}

\section{Introduction}

With the rapid development of IT infrastructure, a massive amount of data has become available, making machine learning-based solutions more feasible and cost-effective. Federated learning, a novel collaborative machine learning approach, addresses data barriers across diverse clients by utilizing more data for training without violating privacy regulations. In a typical client-server model, a group of clients collaborates with an aggregation server to form a "federation" for training a global model that benefits from data owned by different clients. This interactive training process requires clients and the server to frequently communicate in a synchronous manner. During each training round, a client receives the latest global model from the server, trains its local model using its local dataset, and sends model updates (e.g., gradients, model parameters) to the server. The server then generates the next global model version through an aggregation process that takes local model updates as input. Throughout this process, raw data never leaves its owner, providing basic privacy protection for participating clients.

Federated learning, while offering privacy benefits, may still leak sensitive client information due to model inversion and membership inference attacks \cite{nasr2018comprehensive}. To protect local model privacy, existing works employ various methods such as differential privacy (DP) \cite{choudhury2019differential, pan2021fl, zhou2022personalized,he2022differentially} and Local Differential Privacy (LDP) \cite{truex2020ldp}. Privacy-Preserving Federated Learning (PPFL) techniques, including homomorphic encryption (e.g., Paillier encryption) \cite{acar2018survey} and Multi-Party Computation (MPC), have also been explored. However, both homomorphic encryption and MPC-based approaches suffer from lower efficiency compared to the original federated learning, limiting their practical application in real-world projects.

Bonawitz et al. \cite{bonawitz2017practical} proposed Secure Aggregation (SecAgg) for federated learning to address the trade-off between model privacy and aggregation efficiency. The noise used to protect model updates are pairwise negotiated between clients to make sure that the sum of noise added by all clients is zero. Considering clients may drop unexpectedly in any step of the protocol and cause the result incorrect, the server needs to collect the secret shares from online clients and recover the secret for every dropped client. However, its synchronized workflow, which requires devices to wait for the slowest one, leads to accumulated idling time and reduced efficiency compared to the original federated learning approach.
Asynchronous federated learning can have higher utilization of computing power and communication bandwidth for devices, which increases the overall efficiency of the system. However, asynchronous federated learning cannot use the original federated aggregation algorithm, because the updates from clients are not based on the same version of a global model. To handle this problem, clients' models should always have tags to denote the version of the global model. The update of the global model can happen when the clients' updates come, by using the mixing algorithm \cite{xie2019asynchronous}, or when there is a certain number of updates from clients received, by using weighted aggregation with staleness function \cite{nguyen2022federated}. Both methods can modify federated learning into an asynchronous alternative. However, privacy preservation becomes a concern, because mask-based secure aggregation requires synchronization. The aggregation method of one-shot recovery \cite{yang2021lightsecagg} allows clients to communicate with the server without synchronization in the early rounds of one training epoch, but synchronization is still required when one-shot mask removal happens. So et al. \cite{so2021secure} proposed secure aggregation in buffered asynchronous federated learning, in which the server maintains a buffer to temporarily store the coming updates and performs the aggregation and updates the global model when the buffer is full.

In this paper, we study the method to achieve secure aggregation for fully asynchronous federated learning and propose a Privacy-Preserving Asynchronous Federated Learning protocol (PPA-AFL) which facilitates secure aggregation of clients' local models while enhancing the efficiency of global model training. The conflict between local model leakage and local model aggregation is addressed by employing Paillier encryption in asynchronous federated learning, ensuring that clients' updates' cleartext remains concealed while maintaining the same aggregation result. By implementing a dual-server setting and threshold secret sharing, local model aggregation is only executed when a specific number of clients' updates are received, effectively mitigating the risk of global model leakage related to individual clients by reducing the contribution proportion of single clients in each aggregation.
The proposed protocol offers full asynchronicity from the clients' perspective, allowing them to determine their participation and eliminating the requirement for continuous online presence until aggregation. Consequently, the overhead imposed on clients by the system is comparable to that of the original federated learning approach.

\textsf{Outline.} The remainder of this paper is organized as follows: Section \textsc{II} formalize the system model, security threats, and design goals of our proposed solution. The detailed construction of our novel design is presented in Section \textsc{III}, followed by an in-depth security analysis in Section \textsc{IV}. Finally, Section \textsc{V} showcases the performance evaluation of our design, and Section \textsc{VI} offers concluding remarks.

\section{Problem Statement}

We present the entities in the system, the security model, and the goals of the proposed PPA-AFL.

\subsection{Entities}

The dual-server federated learning system consists of three entities: an encryption server, an aggregation server, and clients.

\begin{itemize}
\item
The encryption server: The encryption server is a relatively powerful machine that has high computing ability, and a reliable network connection. The server does not have data for the training task. In the training, this server generates keys for homomorphic encryption and secret shares for threshold aggregation. This server can communicate with clients bidirectionally to allow clients to join the training at any time. The incentive of the server is to receive a commission from the duties of managing the cryptographic system.
\item
The aggregation server: The server is a relatively powerful machine that has high computing ability, reliable network connection, and large storage. The server does not have data for the training task. The incentive of the server is to produce a global model. This server maintains a buffer to store updates from clients and performs the aggregation on the ciphertext of local models. The aggregation result is sent to the other server for decryption.
\item
Clients: A client is a device with the local dataset. It is assumed that a client does not have high computing ability, reliable network connection, and large storage. A single client has limited data that is not diverse enough but the gathering of data from multiple clients can cover throughout data distribution. To find benefits in a global model that has better generalization ability, clients are motivated to join federated training. Clients in this system can occasionally contribute to some rounds of aggregation.
\end{itemize}

\begin{figure}[ht]
\centering
\includegraphics[width = 3.5in]{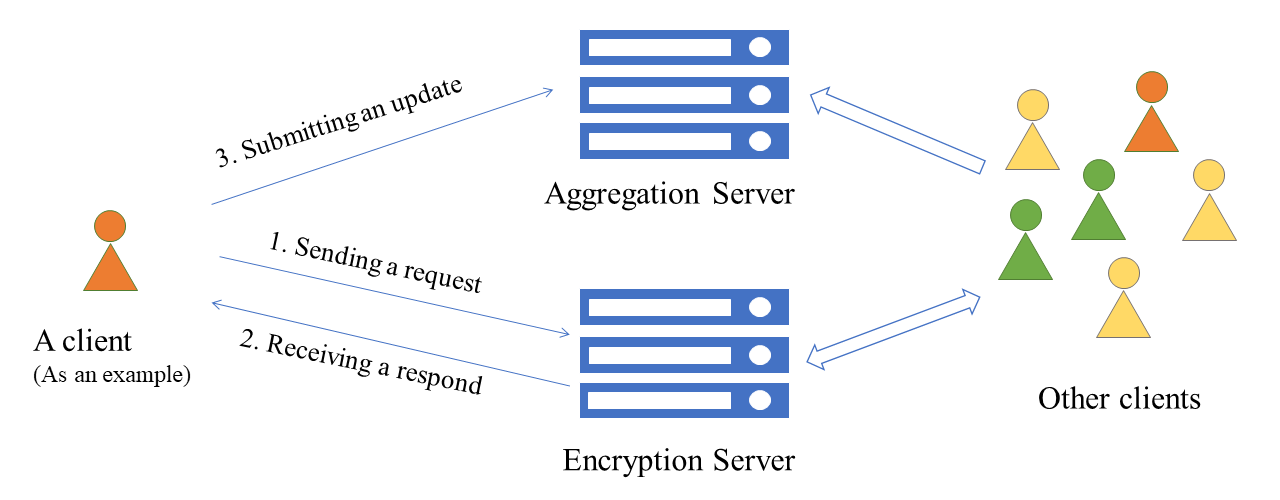}
\caption{System Model for PPA-AFL}
\label{fig:3.1}
\end{figure}

The system model is depicted in Fig.\ref{fig:3.1}. In the original federated learning, there is only one server and multiple clients. In the proposed design, two servers are responsible for different parts of the protocol. The first server is called the aggregation server, which performs the aggregation of updates. The second server is called the encryption server, which distributes keys for encryption and decrypts the final result. The aggregation server maintains a buffer to save the updates from clients. The encryption server keeps the global model of different versions, and maintains the key pairs and secret shares for the next round of aggregation. The clients have their dataset on the local storage, and train the local model based on the global model using local data.

In this system, communication does not happen between arbitrary two parties. Two servers can communicate. Clients and encryption servers can communicate bidirectional, while the communication between clients and aggregation is only from clients to the server.

Clients contribute to the global model by doing local training and submitting updates. The aggregation server performs the aggregation, which uses the local models' ciphertext and outputs the global model's ciphertext. The aggregation server should only expose the encrypted global model to the encryption server but not any individual encrypted update. The encryption server generates keys for homomorphic encryption and sends them to clients who claim they are about to do the local training. The encryption server also decrypts the ciphertext from the aggregation server to get the new global, which is sent to clients along with encryption keys. The encryption server should only decrypt the ciphertext of the global model when the global model is aggregated from a large enough number of clients' updates, which is implemented by using secret sharing.

\subsection{Security Threats}

There are three parties in our protocol, the encryption server, the aggregation server, and the clients. There are two threat models for our system:

\begin{itemize}
\item
All entities in our system are assumed to be honest but curious.
\item
The aggregation server and the encryption server are honest but curious. Some clients are malicious, and the rest of the clients are honest but curious.
\end{itemize}

For every client in the system, it has a local dataset that it is not willing to share. The raw data in the dataset of other clients is not accessible in the federated learning, but the updates that contain the information of the local data can be analyzed to expose the sensitive information. Unprotected updates from clients can become a possible privacy leakage. Additionally, because a federated learning system is open and welcoming clients from the internet in practice, the communication channel may not be secure. The motivation of secure federated learning is to perform the federated learning, while protecting the clients' privacy at the same time.

There are three possible threats to the privacy of clients, when ``all entities in our system are assumed to be honest but curious" assumption is applied:

\begin{itemize}
\item
In our protocol, updates are sent from the clients to the aggregation server. The channel between a certain client and the aggregation server may not be secure, which means an attacker may access the updates of certain clients. The protocol needs to protect clients' privacy, even when their updates are known by an attacker.
\item
Aggregation happens on the aggregation server. The aggregation server performs a linear operation on the ciphertext of the clients' local model. The result is sent to the encryption server to be decrypted. The fewer updates are included in an aggregation, the more possibility of a certain client's privacy can leak. The minimum number of updates that must be reached to allow an aggregation start should be considered.
\item
The encryption server has keys to encrypt, decrypt, and perform the evaluation on the ciphertext. When a certain update is received by the encryption server, the plaintext of that client's local model is exposed. The encryption server in our protocol should never receive a ciphertext of the model that only contains one's or a fewer clients' information.
\end{itemize}

In addition, when some clients in our system are malicious, possible threats to the privacy of clients are:

\begin{itemize}
\item
Malicious clients can use a model in which every parameter is zero as its local model. When there are multiple malicious clients that work together, they can get the sum of other clients' local models. The extreme situation is that, in one aggregation, there is only one honest client and all the rest clients are malicious.
\item
Malicious clients can send requests to the encryption server at a high frequency to get more shares. When the number of shares surpasses the recovery threshold, it can determine when to start the next aggregation. With this advantage, it may exploit more information from honest clients.
\end{itemize}

\subsection{Design Goals}

The main goal is to achieve both privacy protection and efficient asynchronous model aggregation in federated learning. To achieve this goal, the following issues need to be addressed:

\begin{itemize}
\item
Local model privacy: The locally trained models and gradients can still leak the sensitive information of the training sample. To protect the privacy of clients, both their local dataset and local models shall not be available in the form of plaintext for any other parties.
\item
Model aggregation conditions: The aggregated global model contains the information of local models. To reduce the information leakage of a specific sample, the global model must be aggregated from at least a certain number of local models, which means threshold aggregation should be adopted.
\item
The staleness of local models: In asynchronous federated learning, the aggregation server receives local models trained on different versions of global models. In order to ensure the validity of the model, the version of the local model needs to be tracked.
\end{itemize}

\section{Proposed PPA-AFL}

In this section, we review the cryptographic primitives and the detailed construction of PPA-AFL.

\subsection{Cryptographic Primitives}

In the proposed work, secret sharing and homomorphic encryption are adopted.

\subsection{Secret Sharing}

Shamir's Secret Sharing (SS) is one of the widely used secret sharing schemes, in which the generated shares have additive homomorphism. It consists of the following three algorithms:

\begin{itemize}
\item
$SS.Setup(1^k)$:
On inputting the security parameter, the algorithm gives $Param \gets SS.Setup(1^k)$. $Param$ is the parameter for the secret sharing protocol and is implicitly used in the following algorithms.
\item
$SS.Share(n,t,m)$:
There is a pre-determined number of shares $n$ and a recovery threshold $t$, and a message $m$. Call the algorithm to get $\{[s_i]\}_{i \in [0,n)} \gets SS.Share(n, t, m)$. $\{[s_i]\}_{i \in [0,n)}$ are $n$ shares.
\item
$SS.Combine(t,\{[s_i]\}_{i \in P'})$:
There is $\{[s_i]\}_{i \in P} \gets Share(n, t, m)$. To be noted, $P$ is a set of virtual parties to hold the shares and $|P|=n$. $P'$ is a subset of $P$, in additional $|P'| > t$. The combination algorithm outputs the message as $m \gets SS.Combine(t, \{[s_i]\}_{i \in P'})$
\end{itemize}

Paillier encryption (PE) is one of the widely used homomorphic encryption schemes that support additive homomorphism. It consists of the following five algorithms:

\begin{itemize}
\item
$PE.Setup(1^k)$:
On inputting the security parameter, the algorithm gives $Param \gets PE.Setup(1^k)$. $Param$ is the parameter for the Paillier encryption scheme and is implicitly used in the following algorithms.
\item
$PE.Keygen(Param)$:
On inputting $Param$ generated from $PE.Setup(Param)$, the algorithm randomly generates a pair of keys $pk$ and $sk$.
\item
$PE.Enc(m,pk)$:
On inputting the public key $pk$ and a message $m$, the algorithm gives $c \gets PE.Enc(m,pk)$. $c$ is the ciphertext corresponding to the message $m$. The related key pair is $(pk,sk)$.
\item
$PE.Eval(\{[c_i]\},f)$:
On inputting a set of ciphertext ${[c_i]}$ and a linear function $f$, the algorithm gives $c_{eval} \gets PE.Eval(\{[c_i]\},f)$. All ciphertext in $\{[c_i]\}$ is related to the same key pair $(pk,sk)$. This evaluation should show homomorphism, which means the decryption of $c_{eval}$ is as same as the result of feeding the decryption of $\{[c_i]\}$ to the linear function $f$.
\item
$PE.Dec(c,sk)$:
On inputting the secret key $sk$ and a ciphertext $c$, the algorithm gives $m \gets PE.Dec(c,sk)$. $m$ is the plaintext corresponding to the ciphertext $c$.
\end{itemize}

\subsection{The Detailed PPA-AFL}

There are three kinds of parties in our PPA-AFL, the encryption server, the aggregation server, and the clients. During system initialization, the encryption server calls $SS.setup(1^k)$ and $PE.setup(1^k)$ to setup parameters.

\subsubsection{Encryption Server}
This server maintains a global accessible value $tag$, which indicates the version of the variables and messages in this system. At the beginning of the protocol, $tag$ is set to $tag=0$. During training, if the current value $tag$ is $n$, the new $tag$ is assigned as $n+1$ when the tag needs to be updated. The tag is maintained by the encryption server, and the server behaves following the description below.

\begin{itemize}
\item
After the aggregation or when $tag=0$. The current value of $tag$ is $v$.
At the same time, the server randomly generates a secret value $s_v$. The server calls $\{[s_{v,i}]\}_{i \in [0,n)} \gets SS.share(n_v,t_v,s_v)$, where $n_v$ controls the number of shares and $t_v$ controls how many updates are needed to allow the aggregation. $n$ is a large enough integer that should always be greater than the number of updates in the system. $t$ is selected by the encryption server to limit the behaviour of the aggregation server. Then, the server calls $(pk_v, sk_v) \gets PE.keygen()$.
\item
When receiving a request from a client $i$, the server responds to the client a message with the following values: the current tag $v$, the public key $pk_v$, a secret share that has not been sent to others yet $s_{v,x}$, and the latest global model $M_v$.
\item
During the aggregation process, upon receiving the ciphertext of the new global model $c_{v,M}$ and the shares $\{[s_{v,i}]\}_{i \in [0,t')}$ from the aggregation server, the server first verifies if $t' \geq t$, and computes $s_v=SS.combine(t_v,\{[s_{v,i}]\}_{i \in [0,t')})$. If the verification result is satisfactory, the server proceeds to call $M_{v+1} \gets PE.dec(c_{v,M}, sk_v)$ to obtain the updated global model; otherwise, the aggregation is deemed unsuccessful, leaving $tag$ and the global model unchanged. This information should be disseminated to all parties in the system as notification.
\end{itemize}
\subsubsection{Clients}
When the client $i$ wants to join the train, it sends a request to the encryption server. In the response from the server, it gets the current tag $v$, the public key $pk_v$, a secret share $s_{v,x}$, and the latest global model $M_v$. This client runs the local training to get $m_{v,i,count}$. $count$ indicates the number of the local models under the same $v$, i.e., $count^{th}$. This value may be omitted in the following description. The client calls $c_{v,i,count} \gets PE.enc(m_{v,i,count},pk_v)$ to get the ciphertext of the local model. The client sends an update to the aggregation server with the following values: the encrypted local model $c_{v,i,count}$ and a secret share $s_{v,x}$.

To join the training again, the client needs to send a request to the encryption server again. Generally, there is no limit on how many times a client can create updates under the same $tag$.

\subsubsection{Aggregation Server}
The aggregation server maintains a buffer to save the updates from clients temporarily.

When the number of local model updates with the same $tag=v$ surpass $t_v$, the server calls $c_{v,M} \gets PE.eval(\{[c_{v,x}]\},f)$ to obtain the ciphertext of the aggregation result. $f$ is the aggregation algorithm, which is used in the original federated learning.

The shares in these updates are packaged as $\{[s_{v,x}]\}$. The subtext $x$ is used for simplicity, while the local models and shares are still from clients with information like ``client $i$" and ``update times $count$".

The updates with the old $tag$ whose related global model has been decrypted by the encryption server are discarded. For the new updates with a different $tag$, the first one that reaches the number of $t_v$ is processed as above. The message sent to the encryption server contains the corresponding $tag$, the ciphertext of the global model $c_{v,M}$, and a set of shares.

\section{Security Analysis}

In PPL-AFL, the security goals, specifically local model privacy and threshold aggregation, are discussed under two threat models. The first one is all parties are honest but curious. The second threat model is some clients are malicious. When assuming all parties are honest but curious, it is demonstrated that local model privacy is protected, and model aggregation does not occur before the threshold is reached.

A client intending to join federated training first sends a request to the encryption server, receiving a response containing an encryption key generated by the server. The client conducts local training and uses the encryption key to encrypt the local model. All encrypted local models are transmitted to the aggregation server, where aggregation is performed on the ciphertext of local models. The aggregated result is then sent to the encryption server for decryption. Throughout this process, the local model plaintext remains on the local device. Due to the security of the Paillier encryption, without the decryption key, the probability of an adversary distinguishing a ciphertext from a random string in the cipher space is negligible, ensuring the aggregation server and potential eavesdroppers cannot extract information from the ciphertext of local models.

Alongside the encryption key, the encryption server provides a secret share to the client. These shares are submitted to the aggregation server when clients send their updates. To decrypt the global model's ciphertext, the aggregation server must provide these shares to the encryption server. If the secret can be reconstructed from the shares, the encryption server performs decryption; otherwise, the request is denied. Based on the Shamir's secret sharing scheme, secrets cannot be reconstructed without a sufficient number of shares, compelling the aggregation server to execute the threshold aggregation honestly.

If some clients are malicious, malicious clients may request encryption keys and secret shares from the encryption server at a high frequency. To make such requests, a malicious client must expose its identity to the encryption server. By maintaining a record of requests, the encryption server can easily detect this attack, allowing for the suspension of the abnormal client and associated aggregation.

\section{Evaluation}

To evaluate the performance of our proposed protocol, we implement it with java. In this section, the time complexity of different parties is analyzed, and the running time is shown.

\subsection{Complexity Analysis}

For simplicity, the setting is given here.
\begin{itemize}
\item
Global models and local models are of size $m$.
\item
The number of updates with the same $tag$ is $u$.
\item
``$t$ out of $n$" secret shares are generated by the encryption server.
\end{itemize}

We discuss the communication cost and computation cost of our PPA-AFL for different parties.

\begin{itemize}
\item
For clients:
\begin{itemize}
\item
Communication cost\\
$O(m)$ for each update. The cost to send a request to the encryption server is $O(1)$. The cost to receive the response from the encryption server is $O(1)$. The cost to send the update to the aggregation server is $O(m)$.
\item
Computation cost\\
$O(m)$ for each update. The cost to encrypt the plaintext of the local model is $O(m)$.
\end{itemize}
\item
For the encryption server:
\begin{itemize}
\item
Communication cost\\
$O(u)$ for the period between two aggregations. The cost to receive a request from clients is $O(u)$ and the cost to send a response to clients is $O(u)$.

$O(m)$ for each aggregation. The cost to receive a ciphertext of a global model from the aggregation server is $O(m)$.
\item
Computation cost\\
$O(n)$ for the period between two aggregations. The cost to generate ``$t$ out of $n$" shares is $O(n)$.

$O(u^2)$ or $O(m)$ for each aggregation, the larger one should be applied. The cost to check the recovery result from shares is $O(u^2)$, and the cost to decrypt the global model is $O(m)$.
\end{itemize}
\item
For the aggregation server:
\begin{itemize}
\item
Communication cost\\
$O(mu)$ for the period between two aggregations. The cost to receive updates from clients is $O(mu)$.

$O(m)$ for each aggregation. The cost to send a ciphertext of the global model to the encryption server is $O(m)$.
\item
Computation cost\\
$O(mu)$ for each aggregation. The cost to perform the aggregation is $O(mu)$.
\end{itemize}
\end{itemize}

\subsection{Experiment Results}

\begin{figure*}
\centering
\subfigure[Time Cost for Local Model Encryption]{
\label{Inte1}
\includegraphics[width=0.32\textwidth]{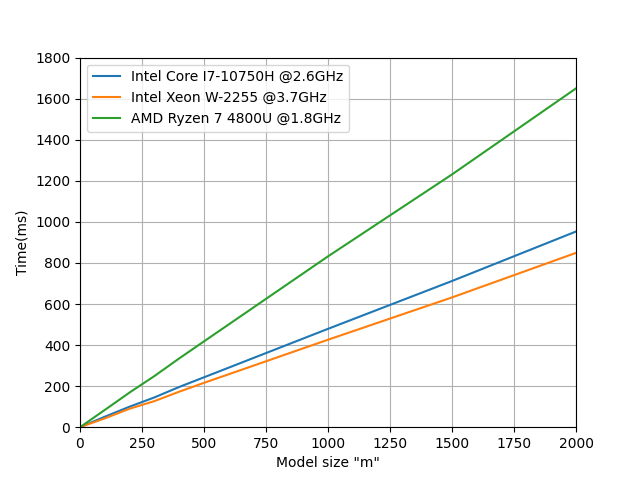}}
\subfigure[Time Cost for Local Model Decryption]{
\label{Inte2}
\includegraphics[width=0.32\textwidth]{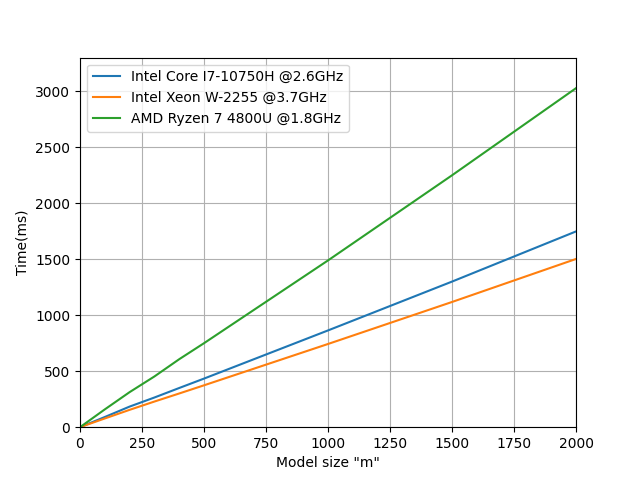}}
\subfigure[Time Cost for Mobile Aggregation with respect to "u"]{
\label{Fig6}
\includegraphics[width=0.32\textwidth]{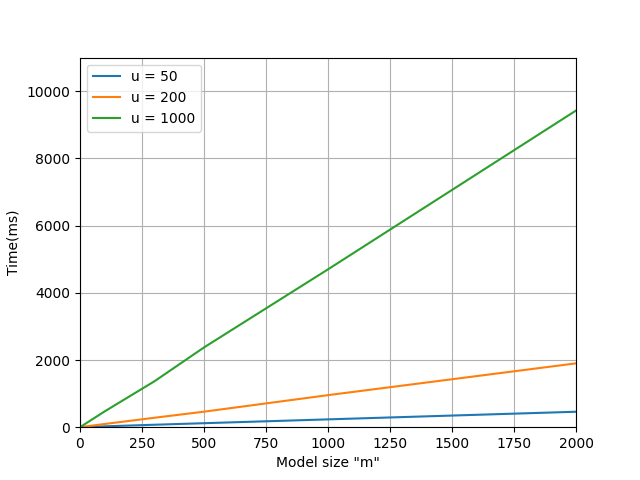}}

\subfigure[Time Cost for Mobile Aggregation with respect to "m"]{
\label{Inte1}
\includegraphics[width=0.32\textwidth]{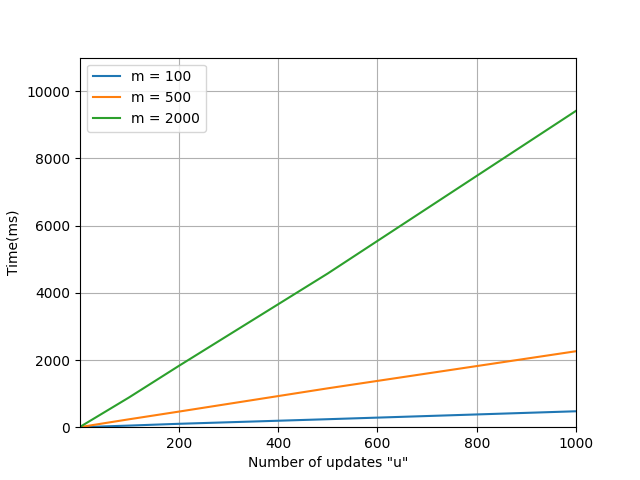}}
\subfigure[Time Cost for Secret Shares Generation]{
\label{Inte2}
\includegraphics[width=0.32\textwidth]{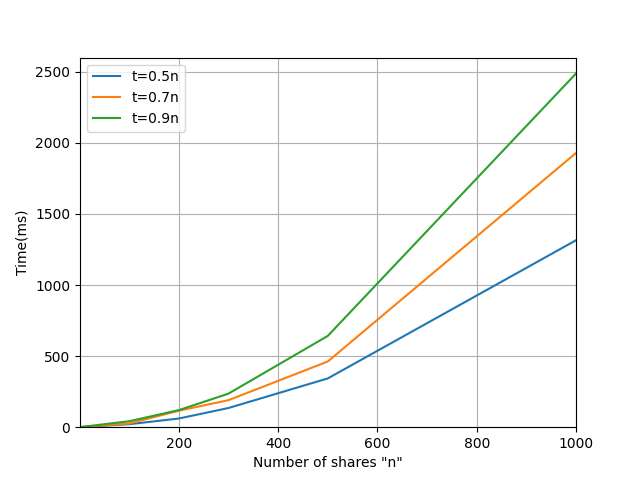}}
\subfigure[Time Cost for Secret Recovery]{
\label{Fig6}
\includegraphics[width=0.32\textwidth]{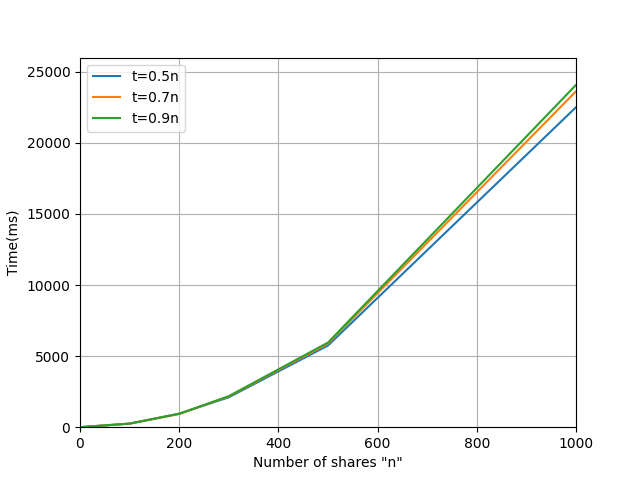}}

\caption{Performance Evaluation of PPA-AFL}

\end{figure*}

We implement our PPA-AFL in java and acquire the running time. The computers have the CPU Intel Core i7-10750H, Intel Xeon W-2255, and AMD Ryzen 7 4800U. The CPU frequency is locked at 2.60GHz, 3.70GHz, and 1.80GHz, respectively. The above-mentioned hardware is selected to simulate the setting of a relatively powerful personal PC, a common business server, and a slim laptop. The operating system is the latest version of Windows 11. All programs only use a single thread to run.

Because the proposed PPA-AFL is asynchronous, most operations can run in parallel. The efficiency of the system is determined by the slowest operation of each party. The result is shown in Fig. 2.

To evaluate the performance of the homomorphic encryption adopted in PPA-AFL, we use three hardware settings mentioned above to run the encryption and decryption algorithm, the time consumption is shown in Fig. 2a and Fig. 2b. From the result, the time consumption is linear to the model size $m$. When assuming the model size is 1000, which is a common value in practice, a slim laptop can perform the encryption in about 800ms, and a business server can perform the decryption in about 750ms. This reflects that the cryptosystem adopted in PPA-AFL can be implemented with high efficiency.

To assess the performance of the homomorphic aggregation, an Intel Xeon W-2255 powered computer is utilized to execute the homomorphic evaluation algorithm. The time consumption results are depicted in Fig. 2c and Fig. 2d. The time consumption is linearly related to both the model size $m$ and the number of updates $u$. Training a complex model with a large $m$ typically necessitates many updates, resulting in a linear increase in time consumption as the model complexity rises. A trade-off between the model size $m$ and the number of updates $u$ should be carefully considered. Within a reasonable range, reducing the number of updates $u$ for each aggregation can decrease time consumption.

To evaluate the performance of the secret sharing scheme implemented in PPA-AFL, the sharing and recovery algorithms are executed on a computer with Intel Xeon W-2255. The time consumption results are illustrated in Fig. 2e and Fig. 2f. These figures indicate that the time consumption for generating and recovering shares is quadratic with respect to the number of shares $n$ and the threshold of recovery $t$. The selection of $n$ is contingent on the maximum number of updates allowed in the aggregation, while the choice of $t$ depends on the minimum number of updates permitted in the aggregation. In practice, adopting a more frequent aggregation with fewer updates can enhance system performance.

\section{Conclusions}
In this paper, we have proposed a fully asynchronous secure federated learning protocol, which mitigates the effects of device heterogeneity during federated training. The proposed protocol enables federated learning to maximize the use of clients' computational resources, facilitating frequent aggregations when other factors remain constant. Asynchronous aggregation also results in reduced impact from communication latency on the system. Collectively, these advantages contribute to the robust performance in practical applications. However, a notable limitation of the proposed protocol is the requirement for two non-colluding servers. This constraint prevents the protocol's implementation in scenarios where only one party can serve as a server or when the parties acting as servers have the motivation to collude.

\bibliographystyle{IEEEtran}
\bibliography{References}

\end{document}